\documentstyle{article}
\begin{document}
\newcommand{\be}{\begin{equation}}
\newcommand{\ee}{\end{equation}}
\newcommand{\ba}{\begin{eqnarray}}
\newcommand{\ea}{\end{eqnarray}}
\newcommand{\bi}{\bibitem}
\def\Res{\mathop{\rm Res}\nolimits}
\def\Re{\mathop{\rm Re}\nolimits}
\def\Im{\mathop{\rm Im}\nolimits}

\begin{titlepage}
\hbox to \hsize{\large \hfil  IHEP 2000-40}
\hbox to \hsize{\hfil hep-ph/0011178}
\hbox to \hsize{\hfil October, 2000}
\vfill
\large \bf
\begin{center}
On  dependence of nonperturbative contributions
in $\bar\alpha_s(q^2)$ on an initial approximation of perturbation 
theory in an analytic approach to QCD 
\footnote{Talk presented at the XV International Workshop on High 
Energy Physics and Quantum Field Theory (QFTHEP'00), 
September 14 -- 20, 2000, Tver, Russia}
\end{center}
\vskip 1cm
\normalsize
\begin{center}
{\bf Aleksey I. Alekseev
\footnote{Electronic address: 
alekseev@mx.ihep.su}}\\
{\small Institute for High Energy Physics,\\
142280 Protvino, Moscow Region, Russia}\\
\end{center}
\vskip 1.5cm
\begin{abstract}
In the  framework of analytic approach to QCD, which has been
recently  intensively
developed,  the dependence of nonperturbative contributions
in a running coupling of strong interaction on initial perturbative
approximation to 3-loop order is studied. The nonperturbative 
contributions are obtained in an explicit form. In the
ultraviolet region they are shown to be represented in the form of
the expansion in the inverse powers of Euclidean momentum squared.
The expansion coefficients are calculated for different numbers
of active quark flavors $n_f$ and for different numbers of loops 
taken into account. For all $n_f$ of interest it is shown that
2-loop order and 3-loop order corrections result in partial 
compensation of 1-loop order leading in the ultraviolet region
nonperturbative contribution.
\end{abstract}
\vskip 1cm
PACS number(s): 12.38.Aw, 12.38.Lg
\vfill
\end{titlepage}

It is widely believed that unphysical singularities of the 
perturbation theory in the infrared region of QCD should be
canceled by the nonperturbative contributions.
The nonperturbative contributions arise quite naturally in an
analytic approach~\cite{SolShirTMF} to QCD.
The so-called "analyticization procedure" is used in this approach.
The main purpose of this 
procedure is to remove nonphysical singularities from  approximate 
(perturbative) expressions for the Green functions of QFT. The idea
of the procedure goes back to Refs.~\cite{Red,Bog} devoted to the
ghost pole problem in QED. The foundation of the procedure is the
principle of summation of imaginary parts of the perturbation theory
terms. Then, the K\"allen -- Lehmann spectral representation  results
in the expressions without nonphysical singularities.
In recent papers~\cite{Shir,Shir1} it is suggested to solve the ghost
pole problem in QCD  demanding the 
$\bar\alpha_s(q^2)$  be analytical in $q^2$ (to compare with 
dispersive approach~\cite{Doksh}). 
As a result, instead of the one-loop expression 
$\bar\alpha^{(1)}_s(q^2)=(4\pi/b_0)/\ln (q^2/\Lambda^2)$
taking into account the leading logarithms and having the ghost
pole at $q^2=\Lambda^2$ ($q^2$ is the Euclidean momentum squared),
one obtains the  expression
\be
\bar\alpha^{(1)}_{an}(q^2)=\frac{4\pi}{b_0}\left[\frac{1}{\ln(q^2/
\Lambda^2)}+\frac{\Lambda^2}{\Lambda^2-q^2}\right].
\label{1}
\ee
Eq.~(\ref{1}) is an analytic function in the complex $q^2$-plane
with a cut along the negative real semiaxis. The pole of the
perturbative running coupling at $q^2=\Lambda^2$ is canceled by the
nonperturbative contribution 
$(\Lambda^2\mid_{g^2\rightarrow 0}$ $\simeq \mu^2\exp
\{-(4\pi)^2/(b_0g^2)\})$ and the value $\bar\alpha^{(1)}_{an}(0)=
4\pi/b_0$ appeared finite and independent of $\Lambda$. 
The most important feature of the "analyticization procedure" 
discovered~\cite{Shir,Shir1} is the stability property of the value
of the 
"analytically improved" running coupling constant at zero with 
respect to high corrections, 
$\bar\alpha^{(1)}_{an}(0)=$	$\bar\alpha^{(2)}_{an}(0)=$
$\bar\alpha^{(3)}_{an}(0)$.
This property provides the high corrections stability of 
$\bar\alpha_{an}(q^2)$
in the whole infrared region.

The 1-loop order nonperturbative contribution in Eq.~(\ref{1}) can be
presented 
as convergent at $q^2>\Lambda^2$ constant signs series in the inverse
powers of the momentum squared. 
For "standard" as well as for the iterative 2-loop perturbative input
the nonperturbative contributions in the analytic
running coupling are calculated explicitly in Ref.~\cite{A}.
In the ultraviolet region the nonperturbative contributions can 
be also represented as a series in inverse powers of momentum 
squared.
In this paper we extract in an explicit form the nonperturbative 
contributions to $\bar\alpha_{an}(q^2)$	 up to the 3-loop order in 
the
analytic approach to QCD, and investigate their ultraviolet behavior.
To handle the singularities originating from the perturbative input,  
we develop here the method which is more general than that of
Ref.~\cite{A}.

According to the definition the analytic running coupling is 
obtained  by the integral
representation
\be
a_{an}(x)=\frac{1}{\pi}\int\limits_0^\infty \frac{d\sigma}{x+\sigma}
\rho(\sigma),
\label{2}
\ee
where the spectral density $\rho(\sigma)=\Im a_{an}(-\sigma-i0)$.
It is seen that dispersively-modified coupling of  
form~(\ref{2}) has an
analytical structure which is consistent with causality.
According to the analytic approach to QCD we adopt that
$\Im a_{an}(-\sigma-i0)=\Im a(-\sigma-i0)$ where $a(x)$ is an
appropriately normalized perturbative running coupling.

The behavior of the QCD running coupling $\alpha_s(\mu^2)$
is defined by the renormalization group equation
\be
\mu\frac{\partial\alpha_s(\mu^2)}{\partial\mu}=
\beta(\alpha_s)=\beta_0\alpha_s^2+\beta_1\alpha_s^3+\beta_2\alpha_s^4
+...,
\label{3}
\ee
where the coefficients
$$
\beta_0=-\frac{1}{2\pi}b_0, \,\,\, b_0=11-\frac{2}{3}n_f,
$$
$$
\beta_1=-\frac{1}{4\pi^2}b_1, \,\,\, b_1=51-\frac{19}{3}n_f,
$$
\be
\beta_2=-\frac{1}{64\pi^3}b_2, \,\,\, b_2=
2857-\frac{5033}{9}n_f +\frac{325}{27}n_f^2.
\label{4}
\ee
The first two coefficients $\beta_0$, $\beta_1$ do not depend on
the renormalization scheme choice.
Here $n_f$ is the number of active quark flavors.
The standard three-loop solution of Eq.~(\ref{3}) 
is written in the form of  expansion in inverse powers of 
logarithms~\cite{Data}
$$
\alpha_s(\mu^2)=\frac{4\pi}{b_0\ln(\mu^2/\Lambda^2)}\left[
1-\frac{2b_1}{b_0^2}\frac{\ln[\ln(\mu^2/\Lambda^2)]}{\ln(\mu^2/
\Lambda^2)}+\right.
$$
\be
+\left.\frac{4b_1^2}{b_0^4\ln^2(\mu^2/\Lambda^2)}\times
\left(\left(\ln[\ln(\mu^2/\Lambda^2)]-\frac{1}{2}\right)^2+
\frac{b_2b_0}{8b_1^2}-\frac{5}{4}\right)\right].
\label{5}
\ee
Let us introduce the function
$a(x)=(b_0/4\pi)\bar\alpha_s(q^2)$, where $x=q^2/\Lambda^2$.
Then, instead of~(\ref{5}) one can write
\be
a(x)=\frac{1}{\ln x}-b\frac{\ln(\ln x)}{\ln^2x}
+b^2\left(\frac{\ln^2(\ln x)}{\ln^3x}-\frac{\ln(\ln x)}{\ln^3x}
+\frac{\kappa}{\ln ^3x}\right),
\label{6}
\ee
where the coefficients $b$ and $\kappa$ are equal to
\be
b=\frac{2b_1}{b^2_0}=\frac{102-\frac{38}{3}n_f}
{(11-\frac{2}{3}n_f)^2},
\,\,\,\,
\kappa=\frac{b_0 b_2}{8b_1^2}-1.
\label{7}
\ee
At $n_f=3$ $b_0=9$, and $b=64/81\simeq 0.7901$, 
$\kappa\simeq 0.4147$.
At $x\simeq 1$   the perturbative running coupling is singular.
At large $x$  the 1-loop term of Eq.~(\ref{6}) defines the 
ultraviolet
behavior of $a(x)$ but 
for small $x$ the behavior of the running coupling depends on the 
approximation we adopt,
and at
$x=1$ there are singularities  of a different analytical structure.
Namely, at $x\simeq 1$ 
\be
a^{(1)}(x)\simeq\frac{1}{x-1},\,\,\,
a^{(2)}(x)\simeq-\frac{b}{(x-1)^2}\ln (x-1),\,\,\,
a^{(3)}(x)\simeq\frac{b^2}{(x-1)^3}\ln^2 (x-1).
\label{8}
\ee
This is not an obstacle for the analytic approach which removes
this nonphysical singularities.
By making the analytic continuation of  Eq.~(\ref{6}) into 
the Minkowski
space $x=-\sigma-i0$, one  obtains
$$
a(-\sigma-i0)=\frac{1}{\ln \sigma-i\pi}-\frac{b}{(\ln \sigma
-i\pi)^2}\ln\left(\ln\sigma-i\pi\right)
+b^2\left\{\frac{\ln^2(\ln\sigma-i\pi)}{(\ln\sigma-i\pi)^3}\right.-
$$
\be
-\left.\frac{\ln(\ln\sigma-i\pi)}{(\ln\sigma-i\pi)^3}+
\frac{\kappa}{(\ln\sigma-i\pi)^3}\right\}.
\label{9}
\ee
Function $a(x)$ in Eq.~(\ref{6}) is regular and real for real $x>1$.
So, to find the spectral density $\rho(\sigma)$ we can use the 
reflection principle
$(a(x))^*=a(x^*)$ where $x$ is considered as a complex
variable. Then
\be
\rho(\sigma)=\frac{1}{2i}\left( a(-\sigma-i0)-a(-\sigma+i0)\right).
\label{10}
\ee
By the change of  variable of the form $\sigma=\exp (t)$,
the analytical expression
is derived from~(\ref{2}), (\ref{9}), (\ref{10}) as follows:
$$
a_{an}(x)=\frac{1}{2\pi i}\int\limits^\infty_{-\infty}
dt \, \frac{e^t}{x+e^t}\times
\left\{\frac{1}{t-i\pi}-\frac{1}{t+i\pi}-
\right.
$$
$$
-b\left[\frac{\ln(t-i\pi)}
{(t-i\pi)^2}-\frac{\ln(t+i\pi)}{(t+i\pi)^2}\right]
+b^2\left[\frac{\ln^2(t-i\pi)}{(t-i\pi)^3}-	\frac{\ln^2(t+i\pi)}
{(t+i\pi)^3}-\right.
$$
\be
\left.\left.
-\frac{\ln(t-i\pi)}{(t-i\pi)^3}+\frac{\ln(t+i\pi)}{(t+i\pi)^3}
+\frac{\kappa}{(t-i\pi)^3}-\frac{\kappa}{(t+i\pi)^3}
\right]\right\}.
\label{11}
\ee
Let us see what the singularities of the integrand of~(\ref{11}) 
in the complex  $t$-plane are.
First of all the integrand has  simple
poles at  $t=\ln x\pm i\pi(1+2n)$,
$n=0,1,2,...$. All the residues of the function  
$\exp(t)/(x+\exp(t))$ 
at these points are equal to unity.
Apart from these 
poles the integrand of~(\ref{11}) 
has simple poles at $t=\pm i\pi$, the third order poles and
logarithmic type branch points  which coincide with
the second order and third order poles. Let us cut the complex
$t$-plane in a  
standard way, $t=\pm i\pi-\lambda$, with $\lambda$ being the real
parameter varying from  $0$ to $\infty$.
The integrand in~(\ref{11}) multiplied by $t$ goes to
zero at $\mid t\mid\rightarrow \infty$. That allows one to append
the integration by the arch of the "infinite" radius without 
affecting the value of the integral. Close the integration contour
$C_1$ in the upper half-plane of the complex variable $t$
excluding the singularity at $t=i\pi$. In this case 
an additional contribution emerges due to the integration along
the sides of the cut and around the singularities at $t=i\pi$.
The corresponding contour we denote as $C_2$.

Let us turn to the integration along  contour $C_1$.
For the integrand of Eq.~(\ref{11}) which we denote as
$F(t)$ the residues at $t=\ln(x)+i\pi(1+2n)$, $n=0,1,2,...$  are
as follows
$$
\Res F(t)\mid _{t=\ln(x)+i\pi(1+2n)}=
\frac{1}{\ln(x)+2\pi in}- \frac{1}{\ln(x)+2\pi i(n+1)}-
$$
$$
-b\left[\frac{\ln(\ln(x)+2\pi in)}{(\ln(x)+2\pi in)^2}-
\frac{\ln(\ln(x)+2\pi i(n+1))}{(\ln(x)+2\pi i(n+1))^2}\right]+
b^2\left[\frac{\ln^2(\ln(x)+2\pi in)}{(\ln(x)+2\pi in)^3}
- \frac{\ln^2(\ln(x)+2\pi i(n+1))}{(\ln(x)+2\pi i(n+1))^3}\right.-
$$
\be
-\left.\frac{\ln(\ln(x)+2\pi in)}{(\ln(x)+2\pi in)^3}+
\frac{\ln(\ln(x)+2\pi i(n+1))}{(\ln(x)+2\pi i(n+1))^3}+
\frac{\kappa}{(\ln(x)+2\pi in)^3}-
\frac{\kappa}{(\ln(x)+2\pi i(n+1))^3}\right].
\label{12}
\ee
By using the residue theorem  
one readily obtains  the  contribution $\Delta(x)$ to 
integral~(\ref{11}) from the integration along contour $C_1$.
It reads
$$
\Delta(x)=\frac{1}{2\pi i}\int\limits_{C_1}F(t)\,dt =
\sum\limits^\infty_{n=0}\Res F\left(t=\ln(x)+i\pi(1+2n)\right)=
$$
\be
=\frac{1}{\ln x}-b\frac{\ln(\ln x)}{\ln^2x}
+b^2\left(\frac{\ln^2(\ln x)}{\ln^3x}-\frac{\ln(\ln x)}{\ln^3x}
+\frac{\kappa}{\ln ^3x}\right).
\label{13}
\ee
We can see that this contribution is exactly equal to  initial
Eq.~(\ref{6}). Therefore, we call it a perturbative part of 
$a_{an}(x)$,  $a^{pt}(x)=\Delta(x)$.
The remaining contribution of the integral along  contour $C_2$
can be naturally  called a nonperturbative part of $a_{an}(x)$,
\be
a_{an}(x)=a^{pt}(x)+a^{npt}_{an}(x).
\label{14}
\ee
Let us turn to the calculation of  $a^{npt}_{an}(x)$. We can omit the
terms of the integrand in Eq.~(\ref{11}) which have no 
singularities at $t=i\pi$. Then, we have
$$
a^{npt}_{an}(x)=\frac{1}{2\pi i}\int\limits_{C_2}
dt \, \frac{e^t}{x+e^t}\times
\left\{\frac{1}{t-i\pi}
-b\frac{\ln(t-i\pi)}
{(t-i\pi)^2}\right.+
$$
\be
+b^2\left.\left[\frac{\ln^2(t-i\pi)}{(t-i\pi)^3}
-\frac{\ln(t-i\pi)}{(t-i\pi)^3}
+\frac{\kappa}{(t-i\pi)^3}
\right]\right\}.
\label{15}
\ee
Let us change the variable $t=z+i\pi$ and introduce the function
\be
f(z)=\frac{1}{1-x\exp(-z)}.
\label{16}
\ee
Then, we can rewrite Eq.~(\ref{15}) in the form
\be
a^{npt}_{an}(x)=\frac{1}{2\pi i}\int\limits_{C} dz \, f(z)
\left\{\frac{1}{z}-b\frac{\ln(z)}{z^2}+b^2\frac{\ln^2(z)}{z^3}-
b^2\frac{\ln(z)}{z^3}+\frac{\kappa b^2}{z^3}\right\}.
\label{17}
\ee
The cut in  the complex $z$-plane goes now from zero to $-\infty$.
Starting from $z=-\infty-i0$  contour $C$ goes 
along the lower side of the cut, then
goes around the origin, 
and next it goes further along the upper side of the cut  to  
$z=-\infty+i0$. Here we consider $x$  as a real  variable, $x>1$.
Then, contour $C$ can be chosen in such a way that it does not
envelop  "superfluous" singularities and the conditions used
in Appendix to find the corresponding integrals be satisfied.
Function~(\ref{16}) with its derivatives decrease exponentially
at $z\rightarrow -\infty$, therefore, we shall omit the boundary 
terms
in formulas given in Appendix. We shall need further the explicit 
expressions for the derivatives of $f(z)$, which read
$$
f'(z)=-\frac{x\exp(-z)}{(1-x\exp(-z))^2},\,\,\,
f''(z)=\frac{x\exp(-z)(1+x\exp(-z))}{(1-x\exp(-z))^3},\,\,\,
$$
\be
f'''(z)=-\frac{x\exp(-z)}{(1-x\exp(-z))^4}\left(1+4x\exp(-z)+
x^2\exp(-2z)\right).
\label{18}
\ee
From  Eq.~(\ref{17}) one can obtain
$$
a^{npt}_{an}(x)=-\frac{1}{2\pi i}\int\limits_{C} dz \,\left\{
f'(z)\ln(z)-bf''(z)\left(\ln(z)+\frac{1}{2}\ln^2(z)\right)\right.+
$$
\be
+\left. b^2f'''(z)\left[\left(1+\frac{1}{2}\kappa\right)\ln(z)+
\frac{1}{2}\ln^2(z)+\frac{1}{6}\ln^3(z)\right]\right\}.
\label{19}
\ee
Taking into account that function $f(z)$ with its derivatives
is regular at real negative semiaxis of $z$,
we can rewrite	  equation~(\ref{19}) in the form
$$
a^{npt}_{an}(x)=-\int\limits_{0}\limits^{-\infty} du \,\left[
f'(u)\Delta_1(u)-bf''(u)\left(\Delta_1(u)+\frac{1}{2}\Delta_2(u)
\right)\right.+
$$
\be
+\left. b^2f'''(u)\left(\left(1+\frac{1}{2}\kappa\right)\Delta_1(u)
+\frac{1}{2}\Delta_2(u)+\frac{1}{6}\Delta_3(u)\right)\right],
\label{20}
\ee
where $u$  is real,	$u<0$ and $\Delta_i(u)$ are discontinuities of 
the 
logarithms
\ba
\Delta_1(u)&=&\frac{1}{2\pi i}\left(\ln(u+i0)-\ln(u-i0)\right)=1,
\nonumber \\
\Delta_2(u)&=&\frac{1}{2\pi i}\left(\ln^2(u+i0)-\ln^2(u-i0)\right)=
2\ln(-u),
\label{21}\\
\Delta_3(u)&=&\frac{1}{2\pi i}\left(\ln^3(u+i0)-\ln^3(u-i0)\right)=
3\ln^2(-u)-\pi^2. \nonumber
\ea
Let us introduce the variable $\sigma=\exp(u)$. From Eqs.~(\ref{18}), 
(\ref{20}), (\ref{21}) we obtain
$$
a^{npt}_{an}(x)=-x\int\limits_{0}\limits^{1} d\sigma \,\left\{
\frac{1}{(x-\sigma)^2}-b\left[1+\ln\left(-\ln(\sigma)\right)\right]
\frac{x+\sigma}{(x-\sigma)^3}\right.+
$$
\be
+\left. b^2\left[1-\frac{\pi^2}{6}+\frac{\kappa}{2}+\ln\left(-\ln
(\sigma)\right)+\frac{1}{2}\ln^2\left(-\ln(\sigma)\right)\right]
\frac{x^2+4x\sigma+\sigma^2}{(x-\sigma)^4}\right\}.
\label{22}
\ee
Integrating the terms of Eq.~(\ref{22}) independent of logarithms
one can obtain
$$
a^{npt}_{an}(x)=\frac{1}{1-x}+b\left\{\frac{x}{(1-x)^2}+
x\int\limits^1_0 d\sigma\, \ln(-\ln(\sigma))\frac{x+\sigma}{(x-
\sigma)^3}\right\}+
$$
\be
+b^2\left\{\left(1+\frac{\kappa}{2}-\frac{\pi^2}{6}\right)
\frac{x(1+x)}{(1-x)^3}-
x\int\limits^1_0d\sigma\, \left[
\ln(-\ln(\sigma))+\frac{1}{2}\ln^2(-\ln(\sigma))\right]
\frac{x^2+4x\sigma+\sigma^2}{(x-
\sigma)^4}\right\}.
\label{23}
\ee
This formula gives the nonperturbative contributions in an explicit
form and is convenient for numerical study.

Let us turn to the ultraviolet behavior of the  nonperturbative 
contributions.  Expanding Eq.~(\ref{23}) in inverse powers of $x$ we 
have
\be
a_{an}^{npt}(x)=\sum\limits_{n=1}\limits^{\infty}\frac{c_n}{x^n},
\label{25}
\ee
where
$$
c_n=-1+bn\left[1+ n\int\limits_{0}\limits^{1}d\sigma\, \sigma^{n-1}
\ln\left(-\ln(\sigma)\right)\right]-
$$
\be
-b^2n^2\left\{1-\frac{\pi^2}{6}+\frac{\kappa}{2}+n\int\limits^1
\limits_0 d\sigma\, \sigma^{n-1}\left[\ln\left(-\ln(\sigma)\right)
+\frac{1}{2}\ln^2\left(-\ln(\sigma)\right)\right]\right\}.
\label{26}
\ee
Making the change of variable $\sigma=\exp(-t)$ and integrating 
\cite{Bateman}, \cite{Prudnikov} over $t$, one can find
\be
\int\limits_{0}\limits^{1}d\sigma\, \sigma^{n-1}
\ln\left(-\ln(\sigma)\right)=\int\limits_{0}\limits^{\infty}dt\,
e^{-nt}\ln(t)=-\frac{1}{n}\left(\ln (n)+\gamma\right),
\label{27}
\ee
\be
\int\limits_{0}\limits^{1}d\sigma\, \sigma^{n-1}
\ln^2\left(-\ln(\sigma)\right)=\int\limits_{0}\limits^{\infty}dt\,
e^{-nt}\ln^2(t)=\frac{1}{n}\left[\left(\ln (n)+\gamma\right)^2
+\frac{\pi^2}{6}\right].
\label{28}
\ee
From Eqs.~(\ref{26}), (\ref{27}), (\ref{28}) we  finally  have
\be
c_n=-1+bn\left(1-\gamma-\ln(n)\right)-\frac{1}{2}b^2n^2\left[
1-\frac{\pi^2}{6}+\kappa+\left(1-\gamma-\ln(n)\right)^2\right].
\label{29}
\ee
Here  $\gamma$ is the Euler constant,  $\gamma\simeq 0.5772$.
We can see from Eq.~(\ref{29})
that power series~(\ref{25}) is uniformly convergent at $x>1$
and its convergence radius is  equal to unity.
To see the details of the coefficients $c_n$ behavior
we give  the results of the numerical study in Table I.
The coefficients $c_n$ depend on $n$, $n_f$ and the number of loops
taken into account. The 1-loop order contribution to $c_n$ equals
$-1$ for all $n$ and $n_f$. The 2-loop order corrections to
$c_n$ decreases monotonously with increasing $n$ for $n_f=0,3,4,5,6$.
In this case the correction to the leading term ($n=1$) is positive 
(it varies from $0.36$ at $n_f=0$ to $0.22$ at $n_f=6$),
whereas for the next terms the corrections are negative.
The 3-loop order corrections to
$c_n$ increases at first and then monotonously decreases with  
increasing $n$    for all $n_f$ we consider.

In the ultraviolet region ($x\gg 1$) the nonperturbative 
contributions 
are determined by the first term of the series~(\ref{25}).
At $n_f \le 4$ the 3-loop order corrections change insignificantly
the 2-loop order results for nonperturbative contributions.
At $n_f=5$ 	the 3-loop order correction is $2.4$ times less than
the 2-loop order correction and at $n_f=6$ it exceeds 
the 2-loop order correction. 
We also note that with increasing $n_f$ the decrease of the 2-loop
corrections is equilibrated to some excess  by the increase
of the 3-loop corrections.

It seems to be  most interesting
that an account for the 2-loop order corrections results in some 
compensation of the 1-loop order leading at large $x$ term of the
form $1/x$ and at 3-loop order compensation remains valid.
We can see from  Table I that at $n_f \ge 3$ the 
3-loop order $|c_1|$ is somewhat less than 	2-loop order $|c_1|$.
However, it apparently does not mean that in an analytic approach
the high-loop corrections lead to the total compensation of 
nonperturbative contributions in the ultraviolet region.

I am indebted to B.A.~Arbuzov, V.A.~Petrov, V.E.~Rochev for 
useful discussions. This work has been supported in part by RFBR
under Grants No.~99-01-00091, No.~98-02-16690.

\begin{center}
{\large \bf Appendix}
\end{center}

 We give here the identities we need in our computations which
 can be obtained by means of an integration by parts. Let function
 $f(z)$ of complex variable $z$ is regular in some domain $D$
 where $z=0 \in D$. Dealing with the singularities of 
 the integrands at the origin of the pole type coinciding
 with the logarithmic type branch points, we cut  domain $D$
 along real negative semiaxis. Then, for any contour $C$ in
 cut domain $\tilde D$, which goes from $z_1\neq 0$ to $z_2 \neq 0$, 
 one can find
$$
\int\limits_{C} \frac{dz}{z} f(z)=-\int\limits_{C} dz\,\ln(z) f'(z)+
\ln(z)f(z)\Big\vert^{z_2}_{z_1},
$$
$$
\int\limits_{C} \frac{dz}{z^3} f(z)=-\frac{1}{2}\int\limits_{C} dz\,
\ln(z) f'''(z)+
\left\{-\frac{1}{2z^2}f(z)-\frac{1}{2z}f'(z)+\frac{1}{2}\ln(z)f''(z)
\right\}\Big\vert^{z_2}_{z_1},
$$
$$
\int\limits_{C} \frac{dz}{z^2}\ln(z) f(z)=-\int\limits_{C} dz\,\left(
\ln(z)+\frac{1}{2}\ln^2(z)\right) f''(z)+
\left\{-\frac{1}{z}f(z)-\frac{\ln(z)}{z}f(z)+\ln(z)f'(z)\right.+
$$
$$
+\left.\frac{1}{2}
\ln^2(z)f'(z)\right\}\Big\vert^{z_2}_{z_1},
$$
$$
\int\limits_{C} \frac{dz}{z^3}\ln(z) f(z)=-\int\limits_{C} dz\,\left(
\frac{3}{4}\ln(z)+\frac{1}{4}\ln^2(z)\right) f'''(z)+
\left\{-\frac{1}{4z^2}f(z)-\frac{3}{4z}f'(z)-\frac{\ln(z)}{2z^2}f(z)
\right.-
$$
$$
-\left.\frac{\ln(z)}{2z}f'(z)+\frac{3\ln(z)}{4}f''(z)+\frac{\ln^2(z)}
{4}f''(z)\right\}\Big\vert^{z_2}_{z_1},
$$
$$
\int\limits_{C} \frac{dz}{z^3}\ln^2(z) f(z)=-\int\limits_{C} dz\,
\left(
\frac{7}{4}\ln(z)+\frac{3}{4}\ln^2(z)+\frac{1}{6}\ln^3(z)\right) 
f'''(z)+
\left\{-\frac{1}{4z^2}f(z)-\frac{7}{4z}f'(z)-\frac{\ln(z)}{2z^2}f(z)
\right.-
$$
$$
-\left.\frac{3\ln(z)}{2z}f'(z)+\frac{7\ln(z)}{4}f''(z)-
\frac{\ln^2(z)}
{2z^2}f(z)-\frac{\ln^2(z)}
{2z}f'(z)+\frac{3\ln^2(z)}
{4}f''(z)+\frac{\ln^3(z)}
{6}f''(z)\right\}\Big\vert^{z_2}_{z_1}.
$$

\begin{table}[p]
\caption{The dependence of $c_n$ on $n$ and $n_f$ for 1-loop, 2-loop 
and 3-loop cases.
}
\begin{center}
\begin{tabular}{r r r r r r r}\hline \hline
     &$n$& $c^{1-loop}_n$  &$\Delta_{2-loop}$  & $\Delta_{3-loop}$ 
	 & $c_n^{2-loop}$ &$c_n^{3-loop}$\\ \hline

$n_f=0$&1  & -1.0& 0.35640 &   -0.01568 &   -0.64360 &   -0.65929 \\
     &2  &-1.0& -0.45582 &    0.08741 &   -1.45582 &   -1.36841 \\
     &3  &-1.0& -1.70912 &   -1.03012 &   -2.70912 &   -3.73924 \\
     &4  &-1.0& -3.24886 &   -4.51236 &   -4.24886 &   -8.76122 \\
     &5  &-1.0& -5.00160 &  -11.31238 &   -6.00160 &  -17.31398 \\
     &6  &-1.0& -6.92407 &  -22.24971 &   -7.92407 &  -30.17378 \\
     &7  &-1.0& -8.98770 &  -38.04599 &   -9.98770 &  -48.03369 \\
     &8  &-1.0&-11.17217 &  -59.34790 &  -12.17217 &  -71.52007 \\
     &9  &-1.0&-13.46228 &  -86.74274 &  -14.46228 & -101.20502 \\
     &10 &-1.0&-15.84626 & -120.76945 &  -16.84626 & -137.61571 \\ 
	 \hline
$n_f=3$&1&-1.0&   0.33405&     0.01608&    -0.66595&    -0.64987 \\
     &2  &-1.0&  -0.42724&     0.19623&    -1.42724&    -1.23101 \\
     &3  &-1.0&-1.60196  &  -0.63626  &  -2.60196  &  -3.23823   \\
     &4  &-1.0&-3.04517  &  -3.48652  &  -4.04517  &  -7.53168   \\
     &5  &-1.0& -4.68801 &   -9.19185 &   -5.68801 &  -14.87987  \\
     &6  &-1.0& -6.48996 &  -18.47225 &   -7.48996 &  -25.96221  \\
     &7  &-1.0& -8.42420 &  -31.96169 &   -9.42420 &  -41.38590  \\
     &8  &-1.0&-10.47171 &  -50.22832 &  -11.47171 &  -61.70003  \\
     &9  &-1.0&-12.61824 &  -73.78810 &  -13.61824 &  -87.40634  \\
     &10 &-1.0&-14.85275 & -103.11452 &  -15.85275 & -118.96726  \\ 
	 \hline
$n_f=4$&1  &-1.0&  0.31252 &    0.04949 &   -0.68748 &   -0.63799  \\
     &2  &-1.0& -0.39970 &    0.31340 &   -1.39970 &   -1.08630  \\
     &3  &-1.0& -1.49872 &   -0.23818 &   -2.49872 &   -2.73690  \\
     &4  &-1.0& -2.84891 &   -2.48499 &   -3.84891 &   -6.33389  \\
     &5  &-1.0& -4.38587 &   -7.15989 &   -5.38587 &  -12.54576  \\
     &6  &-1.0& -6.07168 &  -14.89306 &   -7.07168 &  -21.96474	\\
     &7  &-1.0& -7.88126 &  -26.23939 &   -8.88126 &  -35.12065  \\
     &8  &-1.0& -9.79681 &  -41.69613 &  -10.79681 &  -52.49294  \\
     &9  &-1.0&-11.80500 &  -61.71490 &  -12.80500 &  -74.51990  \\
     &10 &-1.0&-13.89549 &  -86.71011 &  -14.89549 & -101.60561  \\ 
	 \hline
$n_f=5$&1&-1.0&  0.27813 &    0.11653 &   -0.72187 &   -0.60535  \\
     &2  &-1.0& -0.35571 &    0.55755 &   -1.35571 &   -0.79817  \\
     &3  &-1.0& -1.33377 &    0.50736 &   -2.33377 &   -1.82641  \\
     &4  &-1.0& -2.53536 &   -0.73077 &   -3.53536 &   -4.26613  \\
     &5  &-1.0& -3.90317 &   -3.73728 &   -4.90317 &   -8.64045  \\
     &6  &-1.0& -5.40344 &   -9.01126 &   -6.40344 &  -15.41470  \\
     &7  &-1.0& -7.01386 &  -16.99218 &   -8.01386 &  -25.00604  \\
     &8  &-1.0& -8.71859 &  -28.07387 &   -9.71859 &  -37.79246  \\
     &9  &-1.0&-10.50576 &  -42.61401 &  -11.50576 &  -54.11977  \\
     &10 &-1.0&-12.36618 &  -60.94080 &  -13.36618 &  -74.30698  \\ 
	 \hline
$n_f=6$&1&-1.0&  0.22433 &    0.25378 &   -0.77567 &   -0.52189  \\
     &2  &-1.0& -0.28692 &    1.07460 &   -1.28692 &   -0.21231  \\
     &3  &-1.0& -1.07581 &    1.93179 &   -2.07581 &   -0.14402  \\
     &4  &-1.0& -2.04500 &    2.37205 &   -3.04500 &   -0.67296  \\
     &5  &-1.0& -3.14826 &    2.01775 &   -4.14826 &   -2.13052  \\
     &6  &-1.0& -4.35837 &    0.54419 &   -5.35837 &   -4.81418  \\
     &7  &-1.0& -5.65732 &   -2.33454 &   -6.65732 &   -8.99186  \\
     &8  &-1.0& -7.03234 &   -6.87466 &   -8.03234 &  -14.90700  \\
     &9  &-1.0& -8.47386 &  -13.30888 &   -9.47386 &  -22.78274  \\
     &10 &-1.0& -9.97445 &  -21.85073 &  -10.97445 &  -32.82518  \\ 
	 \hline \hline

\end{tabular}
\end{center}
\end{table}

\end{document}